\documentclass[12pt]{article}
\usepackage{amsmath,amssymb,amsfonts,accents,cite}

\usepackage{stmaryrd}

\def\gg{\mathfrak{g}}
\def\veck{ \underaccent{\tilde}{k} }
\def\vect{ \underaccent{\tilde}{t} }

\begin{document}

\begin{center}
{\large\bf A differential operator realisation approach for constructing
Casimir operators of non-semisimple Lie algebras}\\
~~\\

{\large Fahad Alshammari$^{1,2}$,  Phillip S. Isaac$^1$ and Ian Marquette$^1$}\\
~~\\
$^1$ 
School of Mathematics and Physics, The University of Queensland, St Lucia QLD 4072, Australia \\
$^2$ 
Department of Mathematics, Prince Sattam Bin Abdulaziz University, Saudi Arabia. \\
~~\\
{\small Email: fahad.alshammari@uqconnect.edu.au, psi@maths.uq.edu.au, i.marquette@uq.edu.au}
\end{center}

\begin{abstract}
We introduce a search algorithm that utilises differential operator realisations to find polynomial
Casimir operators of Lie algebras. To demonstrate the algorithm, we look at two classes of examples:
(1) the model filiform Lie algebras and (2) the Schr\"odinger Lie algebras. We find that an abstract
form of dimensional analysis assists us in our algorithm, and greatly reduces the complexity of the
problem.
\end{abstract}

\section{Introduction}

It is well established that Casimir operators play an important role in the representation theory of
Lie algebras, the exact analysis of integrable quantum Hamiltonian systems, and the interplay between
the two. For instance, they are of particular importance in the dynamical
symmetry algebra approach whereby an algebraic Hamiltonian can be written in terms of Casimir operators
corresponding to a chain of Lie subalgebras. Among the best known examples 
are the interacting boson model in nuclear physics and the vibron model in
molecular physics \cite{Fi81,De85,Ia87}. The energy spectrum can ultimately be expressed in terms of the
eigenvalues of the Casimir elements on an irreducible submodule (with respect to the symmetry
algebra) of the Hilbert space.

Another class of models in mathematical physics for which Casimir operators have proven to be useful
is the class of quantum superintegrable models. Such models are characterized by a symmetry algebra
for which the Casimir operators enable the algebraic derivation of the spectrum via methods such
as the Daskaloyannis approach \cite{BonDas1999,Das2001}. The most well known examples are the isotropic and anisotropic
harmonic oscillator, the hydrogen atom and the Smorodinsky-Winternitz systems \cite{Foc35, Lou65,
Mi13, Fr66,Ker10} that are connected with semisimple $\mathfrak{su}(n)$ and $\mathfrak{so}(n+1)$ Lie algebras.

The Casimir operators for semisimple Lie algebras have been calculated for all
the classical Lie algebras ( i.e. $A_n$, $B_n$, $C_n$, $D_n$ ) and the exceptional Lie algebras
$E_6$, $E_7$, $E_8$, $F_4$ and $G_2$  (e.g. see  \cite{Be81,Gr64,Ok77,Po66,Pe68,Ra65}). 
However, for non semisimple Lie algebras, the problem of calculating the Casimir operators
can be quite complicated in practice. For certain examples and classes of non semisimple Lie algebras, 
Casimir operators may be constructed, and methods have been presented, for example, in cases of low dimensional Lie
algebras \cite{Sn14} and those with Levi decomposition in terms of
simple Lie algebras with Heisenberg algebras \cite{Qu88}. More generally, the method of virtual
copies presented by Campoamor-Stursberg and Low \cite{Ca09} is able to produce Casimir operators for 
a wide range of Levi decomposable Lie structures, and generalises the approach of \cite{Qu88}. 
It is also known that one can construct
Casimir operators using an infinitesimal method in which the Casimir operators occur as solutions to a system of
partial differential equations \cite{BelBla66,Ab75,Ca02,Sn05}. Such solutions are not necessarily polynomial and
the so-called generalised Casimir operators may be thus determined. 
There has also been work based on the correspondence between irreducible unitary representations of
the Lie group and coadjoint orbits \cite{An12,An14,Ca69,Go13,Ki76}. 

Another difficulty is that the classification problem of non-semisimple Lie algebras is much more
difficult and fewer results are known \cite{Sn14}. Such structures cerainly arise in the context of
physics. One well-studied class of non-semisimple Lie algebras is the class of finite-dimensional conformal Galilei
algebra, indexed by half-integer $\ell$. The smallest instance of $\ell=\frac{1}{2}$
is the kinematic symmetry algebra of the free Schr\"{o}dinger equation, as found by Niederer in 1972 \cite{Ni72}. The
conformal Galilei algebra with higher values of $\ell$ has been studied from both physical and mathematical
points of view, with the works
\cite{Ni72,Ai12,Ai11,Ai13,Ai10,Ai02,An12,An14,Ba68,Ca03,Do97,Ne97,Pe77} providing some important
contributions associated to these algebraic structures. For example, it has been
observed that physical systems having a connection with the finite dimensional $\ell$-conformal Galilei
($\ell>\frac{1}{2}$) algebra are described by Lagrangians or Hamiltonians with higher order
derivatives. Furthermore, there is a connection with this algebra as a symmetry algebra for generalised
oscillator systems such as the Pais-Uhlenbeck oscillator \cite{An14}.

The purpose of this paper is two-fold. Firstly, we produce a functional search algorithm 
that utilises a differential operator realisation to obtain Casimir operators of any Lie algebra.
We also show how one can take advantage of dimensional analysis to increase the applicability and
feasibility of the approach. 
Secondly,  to demonstrate the approach, we apply the algorithm to model filiform algebras
\cite{Ve70} and the Schr\"{o}dinger Lie algebras (also discussed in \cite{Ca05}) corresponding to arbitrary underlying spatial
dimension $d$. This demonstrates the utility of the algorithm in considering families of Lie algebras. 
We note that the Schr\"{o}dinger algebra is of interest in applications in mathematical physics in its own right, although
we do not explore such applications in this paper. For the Schr\"{o}dinger algebra in particular, we
compare our results to those obtained using the approach of \cite{Qu88,Ca09}. 

\section{ The search algorithm }

Consider a finite-dimensional Lie algebra $\gg$ of dimension $N$ over the real or complex field. A {\em Casimir
operator} of $\gg$ is an element $K\in U(\gg)$ that satisfies $[K,X]=0$ for all $X\in \gg$, where
$[X,Y]=XY-YX$. In other words, a Casimir operator is an element of the centre of $U(\gg)$. It is
important to note that we only consider polynomial Casimir operators. 

Let $\gamma=\{ X_1,X_2,\ldots,X_N \}$ be a basis for $\gg$. It is well known that the PBW theorem provides a
basis $\beta$ of $U(\gg)$ in the form
\begin{equation}
\beta = \left\{ X_1^{\omega_1}X_2^{\omega_2}\ldots X_N^{\omega_N}\ | \
\omega_1,\omega_2,\ldots,\omega_N\geq 0\right\}.
\label{PBWbasis}
\end{equation}
For convenience, we define $\beta_m \subset \beta$ as the set containing monomials in the
generators $X_i$ up to degree $m$, i.e.
\begin{equation}
\beta_m = \left\{ X_1^{\omega_1}X_2^{\omega_2}\ldots X_N^{\omega_N}\ | \
\omega_1+\omega_2+\ldots+\omega_N\leq m\right\}. \label{mbasis}
\end{equation}
A Casimir operator is said to be of {\em order $m$} if it can be expressed as a linear combination
of elements of $\beta_m$.

Our starting point is to implement a direct and naive approach to obtain the polynomial Casimir operators of a
class of non-semisimple Lie algebras. The following algorithm produces a polynomial Casimir operator
$K$ of order $m$, if one exists. 

\begin{enumerate}
\item[(A)] 
Choose an integer $m\geq 2$, the order of $K$. Note that a Casimir operator of order 1 is
a central element of the Lie algebra, and will already be known.
\item[(B)] 
Construct the set $\beta_m$ described in (\ref{mbasis}) above.
\item[(C)] 
Set an operator $K_m$ of order $m$ as
\begin{equation}
K_m=\sum_{\sigma \in \beta_m} f_{\sigma}\sigma, \label{k1}
\end{equation}
where the coefficients $f_\sigma$ are in the underlying field. 
\item[(D)] 
Determine a differential operator realisation $\varrho$ of the basis $\gamma$, in terms of
variables $\{x_1,x_2,\ldots,x_n\}$, of the form
\begin{eqnarray*}
\varrho(X_ {i})= \sum\limits_{j=1}^n  f_{ij}\frac{\partial}{\partial x_{j}}+f_{i0}, \ \   
i=1,2,\ldots,N,
\end{eqnarray*}
where $f_{ij}$  and $f_{i0}$ are  polynomials of $x_{1},x_2,\ldots, x_{n}$. Note that $n$ may be
independent of $N$,$m$. Crucially, $\varrho$ is extended to the enveloping algebra via the
homomorphism property.
\item[(E)] 
Construct  operators  $[\varrho(K_m),\varrho(X_{i})]$ and apply the commutator to an arbitrary
differentiable function $\psi(x_{1},x_2,\ldots, x_{n})$. 
\item[(F)]
Setting $[\varrho(K_m),\varrho(X_ {i})]\psi(x_{1}, ..., x_{n})=0$ gives equations of the form
$$
\sum_{\ell=0}^m\sum_{\veck\in J_\ell}\sum_{\vect\in I_{\veck}}\sum_{\sigma\in\beta_m}
\omega_{i\sigma \vect\veck}f_\sigma x_1^{t_1} x_2^{t_2}\cdots x_n^{t_n}\frac{\partial^\ell
\psi}{\partial x_1^{k_1}\partial x_2^{k_2}\cdots \partial x_n^{k_n}} = 0
$$
where $n$ is the number of variables associated with the realisation, $J_\ell$ denotes the set of
compositions (i.e. ordered partitions) of the integer $\ell$, with $(k_1,k_2,\ldots,k_n) =\veck\in
J_\ell$, $I_{\veck}$ denotes an index set of admissible $(t_1,t_2,\ldots,t_n)=\vect$ (i.e. those
that occur in the expression, depending on the realisation $\varrho$ and the Lie algebra $\gg$), and 
the $\omega_{i\sigma \vect\veck}$ are coefficients in the underlying field.
This leads to a set of linear algebraic equations in the coefficients $f_{\sigma}$ of $K_m$ (from
(\ref {k1}) above).
\item[(G)] Solve the linear algebraic equations 
$$
\sum_{\sigma\in\beta_m}\omega_{i\sigma \vect\veck}f_\sigma =0
$$
for $f_{\sigma}$. In general this may produce a list of $L$ candidate Casimir operators 
$$
\{ K_m^{(1)},K_m^{(2)},\ldots,K_m^{(L)}\}. 
$$
The integer $L$ depends on the Lie algebra $\gg$ and the realisation
$\varrho$ used in previous steps.
\item[(H)] Eliminate any spurious candidate Casimir operators that arise as an artefact of the realisation
employed, or that are functionally dependent on lower order Casimir operators.
\item[(I)] Set 
$$
K_m = \sum_{j=1}^L a_j K_m^{(j)},
$$
and solve for the coefficients $a_j$ by forcing $[K_m,X_{i}]=0$ for each $i=1,2,\ldots N$.
\end{enumerate}

In general, one of the main advantages of this direct approach for constructing Casimir operators is
that it can be applied to any Lie algebra. Moreover, the problem of determining
polynomial invariants can be reduced to that of solving a system of linear algebraic equations. Even for low dimensional
and straightforward cases, however, there are many unexplained subtleties which require further
discussion. We shall elaborate further in the following sections.

\subsection{Differential operator realisations of Lie algebras}

Our approach relies on the existence of a differential operator realisation of the Lie algebra, and
its use in step (D) of the algorithm. For finite dimensional Lie algebras, various constructions
have been proposed to obtain explicitly such operator realisations that are in a usable form for our purposes
\cite{Kos74,Mi68,Do88,Kam90,Dr03,Kli1}.

One class of realisation which is connected with the PDE approach of obtaining Casimir operators is
based on the $\mathfrak{gl}(n)$ embedding. The construction of the realisation is as follows.
Let $\{X_1,\ldots, X_n\}$ be the basis of $\mathfrak g$ with $C^k_{ij}$ the structure constants. The
basis for the coadjoint representation of $\mathfrak g$ in the space $C^{\infty}(\mathfrak g^*)$ is
given by
\begin{eqnarray}
\Hat{X_i}=C^k_{ij} x_k \,   \frac{\partial}{\partial{x_j}}  \label{th22}
\end{eqnarray}
where $\mathfrak g^*$ is the dual space of the vector space $\mathfrak g$.
Note that we adopt the convention of summation over repeated indices.

An advantage of the approach introduced in this
paper is that it does not need to use that particular realisation. This freedom can be exploited as it is
known that even one of the most well-studied Lie algebras, $\mathfrak{sl}(2)$, possesses many
different differential operator realisations which relate to different representations \cite{Ho92,Mi68,Kli1}.
This is also associated to the Askey scheme of orthogonal polynomials \cite{Ask85}. 
Realisations can be of first or second order with some depending on any number of variables from one
up to $N$.
For a Lie algebra in general, different realisations may involve differential operators of different
orders (i.e. not just linear). They also do not necessarily involve the same number of variables,
which can offer advantages from a computational point of view. 
In the case the Lie algebra contains a central extension which, for example, is the case of the Schr\"{o}dinger
algebra \cite{Ni72}, one may rely on a canonical construction \cite{Do88}.

\subsection{A note on dimensional analysis}

The algorithm presented above was merely a starting point to explain a naive approach to computing
Casimir operators. Given the freedom to choose the order of the Casimir operator in step (A), the
algorithm presented above is not at all efficient, and would grow exponentially with the dimension of the
algebra. 
We now seek to modify some of the early steps in the above algorithm in order to improve its
efficiency and make it more palatable for implementation.

At the core of our modified algorithm is a generalised notion of roots at the level of the universal
enveloping algebra, which we refer to as {\em dimensional analysis} in the context of the current
paper, motivated by the use of
the differential operator realisation as discussed in the previous section.

Given a differential operator realisation of the Lie algebra that acts on some space of differentiable
functions, we are able to determine how each generator affects the dimensions (i.e. units) of an
expression. Consequently, we may associate to each algebraic
generator a relative dimension. A similar idea was
discussed in \cite{Ca81} and is of particular use in our current work. As we shall see,
employing dimensional analysis will greatly reduce the amount of calculation involved in determining
the polynomial Casimir operators. The outcome is to express the {\em relative dimension}
of each abstract generator in terms of some suitable artificial dimension. Throughout the text, in
this context we refer to the {artificial relative dimension} of a generator.

To be specific, let $X,Y,Z$ be elements of a Lie algebra such that the Lie product is $[X,Y]=Z,$ 
with $Z$ necessarily being non-zero. 
The artificial relative dimensions of these
elements, denoted $[X],[Y],[Z]$ respectively, must then satisfy $[X][Y]=[Z].$ In other words, the
dimensions must be consistent with the bracket. This definition allows us to
assign a relative dimension to each basis element of the Lie algebra. 

It is clear that this concept of relative dimension is related to that of a root from the
theory of simple Lie algebras. For instance, if the Lie algebra has a Levi decomposition, 
the artificial relative dimensions could be chosen to be consistent with
the roots of the semisimple Levi factor. 
For our purposes, however, it is unnecessary to further develop this connection.

As an example, consider the 4-dimensional nilpotent Lie algebra denoted $n_{4,1}$ (using the
notation of \cite{Sn14})
with basis $\{e_1,e_2,e_3,e_4 \}$ and Lie bracket given by 
$$
[e_2,e_4]=e_1,\ \ \ [e_3,e_4]=e_2,
$$
with all other brackets being zero. The equations for the relative dimensions are then simply 
$[e_2][e_4]=[e_1]$ and $[e_3][e_4]=[e_2].$ In this example, the relative dimensions can all be
expressed in terms of $[e_3]$ and $[e_4]$. Namely, $[e_2]=[e_3][e_4]$ and $[e_1]=[e_3][e_4]^2$.
Alternatively, we may express all of these in terms of artificial dimensions $a$ and $b$
(in this case) by
$$
[e_1]=ab^2,\ \ [e_2]=ab,\ \ [e_3]=a,\ \ [e_4]=b,
$$
or, for this example, simply in terms of the powers of $a$ and $b$ as follows:
$$
[e_1]=(1,2),\ \ [e_2]=(1,1),\ \ [e_3]=(1,0),\ \ [e_4]=(0,1).
$$

\subsection{The modified search algorithm}

We modify the algorithm in the following way to incorporate the use of dimensional analysis
introduced in the previous section. As before, we let $\gamma=\{X_1,X_2,\ldots,X_N\}$ be a basis for
$\gg$. We note that many of the steps below are essentially the same as
their counterparts in the naive algorithm. They have been restated here for completeness.

\begin{enumerate}
\item[(A')] Choose an integer $m\geq 2$, the order of $K$. Note that a Casimir operator of order 1 is
a central element of the Lie algebra, and will already be known. Then
determine the relative dimension $[X_{i}]$ of each basis vector $X_{i} \in \gamma,$ and
all monomials up to degree $m$ in the $X_i$. Choose one relative dimension, say $W$, that occurs
at degree $m$, which should in general occur more than once across all degrees up to $m$. 
This will be the relative dimension of $K$ for which we search. Note that such a $K$ does not
necessarily exist.
\item[(B')]
Construct the set $\beta_m^W\subset \beta$ (with $\beta$ the PBW basis given by (\ref{PBWbasis})),
where
\begin{equation*}
\beta_m^W=\{ X^{\omega_1}_1 X^{\omega_2}_2 ... \ X^{\omega_N}_N|{ \omega_1+ \omega_2+ ... + \omega_N}\leq
m,  \prod\limits_{i=1}^N [X_i]^{\omega_i}=W \}.  
\end{equation*}
\item[(C')] 
Set an operator $K_m^W$ of order $m$ as
\begin{equation}
K_m^W=\sum_{\sigma \in \beta_m^W} f_{\sigma}\sigma, \label{k1dash}
\end{equation}
where the coefficients $f_\sigma$ are in the underlying field. 
\item[(D')]
Determine a differential operator realisation $\varrho$ of the basis $\gamma$, in terms of
variables $\{x_1,x_2,\ldots,x_n\}$, of the form
\begin{eqnarray*}
\varrho(X_ {i})= \sum\limits_{j=1}^n  f_{ij}\frac{\partial}{\partial x_{j}}+f_{i0}, \ \   
i=1,2,\ldots,N,
\end{eqnarray*}
where $f_{ij}$  and $f_{i0}$ are  polynomials of $x_{1},x_2,\ldots, x_{n}$. Note that $n$ may be
independent of $N$,$m$. Crucially, $\varrho$ is extended to the enveloping algebra via the
homomorphism property.
\item[(E')] 
Construct operators $[\varrho(K_m^W),\varrho(X_{i})]$ and apply the commutator to an arbitrary
differentiable function $\psi(x_{1},x_2,\ldots, x_{n})$. 
\item[(F')]
Setting $[\varrho(K_m^W),\varrho(X_{i})]\psi(x_{1}, ..., x_{n})=0$ gives equations of the form
$$
\sum_{\ell=0}^m\sum_{\veck\in J_\ell}\sum_{\vect\in I_{\veck}}\sum_{\sigma\in\beta_m^W}
\omega_{i\sigma \vect\veck}f_\sigma x_1^{t_1} x_2^{t_2}\cdots x_n^{t_n}\frac{\partial^\ell
\psi}{\partial x_1^{k_1}\partial x_2^{k_2}\cdots \partial x_n^{k_n}} = 0
$$
where $n$ is the number of variables associated with the realisation, $J_\ell$ denotes the set of
compositions (i.e. ordered partitions) of the integer $\ell$, with $(k_1,k_2,\ldots,k_n) =\veck\in
J_\ell$, $I_{\veck}$ denotes an index set of admissible $(t_1,t_2,\ldots,t_n)=\vect$ (i.e. those
that occur in the expression, depending on the realisation $\varrho$ and the Lie algebra $\gg$), and 
the $\omega_{i\sigma \vect\veck}$ are coefficients in the underlying field.
This leads to a set of linear algebraic equations in the coefficients $f_{\sigma}$ of $K_m^W$ (from
(\ref{k1dash}) above).
\item[(G')] Solve the linear algebraic equations 
$$
\sum_{\sigma\in\beta_m^W}\omega_{i\sigma \vect\veck}f_\sigma =0
$$
for $f_{\sigma}$. In general this may produce a list of $L$ candidate Casimir operators 
$$
\{ K_m^{W(1)},K_m^{W(2)},\ldots,K_m^{W(L)}\}.
$$
The integer $L$ depends on the Lie algebra $\gg$ and the realisation
$\varrho$ used in previous steps.
\item[(H')] Eliminate any spurious candidate Casimir operators that arise as an artefact of the realisation
employed, or that are functionally dependent on lower order Casimir operators.
\item[(I')] Set 
$$
K_m^W = \sum_{j=1}^L a_j K_m^{W(j)},
$$
and solve for the coefficients $a_j$ by forcing $[K_m^W,X_{i}]=0$ for each $i=1,2,\ldots N$.
\end{enumerate}

In this modified algorithm, the number of terms to consider is greatly reduced, along with the
number of spurious Casimir operators.


\section{Examples}

In order to demonstrate the algorithm at work, we shall apply it to two families of Lie algebras.
Through these examples we attempt to clarify some of the details of our algorithm. In both cases, we
work over the field $\mathbb{C}$.

The first is the class of so-called model filiform Lie algebras. The filiform Lie algebras are defined to be
maximally nilpotent Lie algebras \cite{Ve70,BoFeNu01,CeNuTe17,GoJiKh98,Ha91,GoKh96}. 
The model subclass is the basic family from which all other filiform Lie algebras can be 
described as linear deformations. 
The model filiform Lie algebras are known to have a number of Casimir operators, and so provide a
suitable source of examples for our purposes.

The second example is the class of Schr\"{o}dinger algebras, i.e. the kinematic symmetry algebra of the
free Schr\"{o}dinger equation in $d$ spatial dimensions. While the number of Casimir operators grows
with the rank of the Levi factor (semisimple part), we demonstrate how to use our algorithm to
produce quadratic and cubic Casimir operators, in arbitrary dimensions.

\subsection{Model filiform Lie algebras}

The model filiform algebra, denoted $L_n$, of dimension $n\geq 3$ is a nilpotent Lie algebra with basis
$\{e_1,e_2,\ldots,e_n\}$ and non-zero Lie bracket given by
$$
[e_k,e_n]=e_{k-1},\ \ k=2,3,\ldots,n-1.
$$
Clearly $e_1$ is a Casimir operator. Here we work through our algorithm to produce the higher order
Casimir operators. 

We first remark that the commutator table, denoted ${\cal C}(L_n)$, is as follows:

\begin{table}[h!]
  \centering
  \begin{tabular}{|l|l|l|l|l|c|r|}
    \hline
 [ , ]& $e_{1}$ & $e_{2}$& $e_{3}$& $...$& $e_{n-1}$& $e_{n}$\\
    \hline
 $e_{1}$ & 0 & 0 & 0 &  & 0 & 0\\
  \hline
 $e_{2}$ & 0 & 0 & 0 &  & 0 & $e_{1}$\\
  \hline
 $\vdots$ &  &  &  &  &  &$\vdots$\\
  \hline
 $e_{k}$ & 0 & 0 & 0 &  & 0 &$e_{k-1}$\\
  \hline
 $\vdots$ &  &  &  &  &  &$\vdots$\\
  \hline
 $e_{n-1}$ & 0 & 0 & 0 &  & 0 &$e_{n-2}$\\
  \hline
  $e_{n}$ & 0 & -$e_{1}$ & -$e_{2}$ & $\cdots$ & -$e_{n-2}$& 0\\
  \hline
  \end{tabular}
\end{table}

Regarding the number of Casimir operators for $L_n$, the Beltrametti-Blasi formula \cite{BelBla66}
gives the value $n-\mbox{rank}(\overline{\cal C}(L_n))$, where $\overline{\cal C}(L_n)$ denotes the
commutator table treated as a numerical matrix with generic non-zero entries. In this case, it is clear that rank$(\overline{\cal
C}(L_n))=2,$ so the number of (generalised) Casimir operators is $n-2$. It turns out that we find
precisely this many independent polynomial invariants, one of each degree up to $n-2$. 

Following the algorithm, we need to first establish the relative dimensions of the basis elements.
This is easily achieved by setting
$$
[e_k]=[e_{n-1}][e_n]^{n-k-1}, \ \ k=1,\ldots,n-2
$$
For convenience, we simplify this labelling so that we only use the exponents, and write
$$
[e_k]=(1,n-k-1),\ \ k=1,\ldots,n-1,\ \ [e_n]=(0,1).
$$
Now for terms of degree 2, we observe that the quadratic terms of equal relative dimensions occur
in the following sets (using the notation of step (B') in the modified algorithm):
\begin{align*}
\beta_2^{(2,2n-2k-4)} &= \left\{e_{k+2-\ell}e_{k+\ell}\ |\ \ell=1,2,\ldots,\left\lfloor\frac{n}{2}\right\rfloor -
\left\lfloor \left| \frac{n}{2}-k-1\right| \right\rfloor \right\},\ \ k=1,2,\ldots,n-3, \\
\beta_2^{(2,2n-2k-5)} &= \left\{e_{k+2-\ell}e_{k+1\ell}\ |\ \ell=1,2,\ldots,\left\lfloor\frac{n-1}{2}\right\rfloor -
\left\lfloor \left| \frac{n-1}{2}-k-1\right| \right\rfloor \right\},\ \ k=1,2,\ldots,n-4.
\end{align*}
It is worth at this stage pointing out a subtlety. That is, in the statement of the algorithm, we specify that
we should construct the sets to contain terms of degree {\em up to} $m$. In this example, however,
due to the nature of the Lie bracket, we will only find homogeneous terms of a particular degree. We
also note that the PBW basis is based on the natural ordering of generators: $e_1,e_2,\ldots,e_n$.

For this example, we work with the vector field realisation \cite{Sn14}
$$
e_1=0,\ \ e_k = x_{k-1}\frac{\partial}{\partial x_n},\ \ 1<k<n,\ \
e_n=-\sum_{k=2}^{n-1}x_{k-1}\frac{\partial}{\partial x_k}.
$$
Without including all the tedious details of the calculation, we find that by proceeding through the
algorithm, the outcome is that the sets $\beta_2^{(2,2n-2k-4)}$ produce
functionally independent Casimir operators only for
$k=1,2,\ldots,\left\lfloor\frac{n}{2}\right\rfloor-1$, and the sets $\beta_2^{(2,2n-2k-5)}$ produce
none. The quadratic Casimir operators in this case are
$$
Q_k = e_{k+1}^2 +2 \sum_{j=1}^{k}(-1)^je_{k+1-j}e_{k+1+j},\ \
k=1,2,\ldots,\left\lfloor\frac{n}{2}\right\rfloor-1.
$$
We make the observation that the Lie bracket $[\langle \beta_2^{(2,2n-2k-4)} \rangle,e_n]\subseteq
\langle\beta_2^{(2,2n-2(k-1)-5)}\rangle,$ where we use the notation $\langle\cdot\rangle$ to denote
the span. This is helpful since $|\beta_2^{(2,2n-2k-4)}|=1+|\beta_2^{(2,2n-2(k-1)-5)}|$ when
$k=1,2,\ldots,\left\lfloor\frac{n}{2}\right\rfloor-1$, and so the number of linear equations to solve
is always one less than the number of coefficients specified in the Casimir operator in step (C')
of the algorithm. This is not a guarantee that a non-trivial solution exists, 
but certainly helps the cause in finding a non-trivial
nullspace for the linear equations of step (G').

For completeness, we also give the set of functionally independent cubic Casimir operators as
follows that is output from the algorithm:
\begin{align*}
C_k & = (-1)^k\sum_{j=1}^{k+1}(2k+3-2j)(-1)^je_je_{2k+3-j} + 2(-1)^ke_1e_2e_{2k+1} - e_2e_{k+1}^2\\
 & \qquad +
(-1)^k\sum_{j=1}^k2(-1)^je_2e_{1+j}e_{2k+1-j},\ \
k=1,2,\ldots,\left\lfloor\frac{n+1}{2}\right\rfloor-2.
\end{align*}
The relative dimension is given by 
$$
[C_k] = (3,3n-2k-7).
$$ 
We can see that for $L_n$, the quadratic Casimir operators $Q_k$ and the cubic Casimir operators
$C_k$ are enough to provide a full set of functionally independent Casimir operators. Including
$e_1$, there are indeed $n-2$ of them. For low dimensional cases, the following table summarises the
number of quadratic and cubic Casimirs of $L_n$ in the form given.
\begin{table}[h!]
  \centering
  \begin{tabular}{|c|c|c|c|c|c|c|c|c|c|c|c|c|c|}
    \hline
 $n$& 3 & 4 & 5 & 6 & 7 & 8 & 9 & 10 & 11 & 12 & 13 & $\cdots$ & $n$ \\
    \hline
 \# quad.& 0 & 1 & 1 & 2 & 2 & 3 & 3 & 4 & 4 & 5 & 5 & $\cdots$ &
$\left\lfloor\frac{n}{2}\right\rfloor-1$ \\
  \hline
 \# cub.& 0 & 0 & 1 & 1 & 2 & 2 & 3 & 3 & 4 & 4 & 5 & $\cdots$ &
$\left\lfloor\frac{n+1}{2}\right\rfloor-2$ \\
  \hline
  \end{tabular}
\end{table}

It is well known in the literature \cite{Sn14,Sn05,GoKh96,NdWi94} that an alternative set of functionally independent Casimir
operators are given by  
$$
\xi_k = \frac{(-1)^k k}{(k+1)!} e^{k+1}_{2}+\sum\limits_{i=0}^{k-1} \frac{(-1)^i e^{i}_{2} e_{k+2-i}
e^{k-i}_{1}}{i!},  \ \   1\leq k\leq n-3.
$$
We note, for example, that $Q_1$ is proportional to $\xi_1$, $C_1$ is proportional to $\xi_2$, and
\begin{align*}
8\xi_3 &= 4e_1^2Q_2-Q_1^2,\\
30\xi_4 &= C_1Q_1-6e_1^2C_2,\\
144\xi_5 &= 36e_1^2Q_1Q_2 + 8C_1^2-9Q_1^3-72e_1^4Q_3,
\end{align*} 
and so on, in $L_n$ (here $n\geq 8$ in order to define $\xi_5$ for example). 
This demonstrates in principle how the functionally independent Casimir operators $Q_k$
and $C_k$ may relate to an already known functionally independent set $\xi_k$. We remark that our
methods also produce many higher order Casimir operators, but it is clear that these will all be
functionally related to those already given.


\subsection{Schr\"{o}dinger algebra}

In this section we apply the algorithm to the centrally extended Schr\"{o}dinger Lie algebra 
\cite{Ni72,Pe77,Do97,Hag72,BarXu81}, denoted $\mathfrak{sch}{(d)}$, associated with an underlying
$(d+1)$-dimensional space-time. A method for determining the Casimir operators of this family of Lie
algebras was investigated by Campoamor-Stursberg in \cite{Ca05}, and explicit results were given for
$d=2,3$. Even earlier, the same author provided valubale insights for Lie algebras with the same semisimple 
Levi factors as $\mathfrak{sch}(d)$ were investigated in \cite{Ca03}.

The generators of $\mathfrak{sch}(d)$ are denoted
\begin{equation}
\left\{ M, P_{n,i}, H, D, C, J_{jk}\ |\ n = 0,1,\ 1\leq i\leq d,\  1 \leq j < k \leq d\right\},
\label{Bas}
\end{equation}
and satisfy the following non-trivial commutation relations: 
\begin{eqnarray}
[D,H]&=& 2H,   \,\,[D,C]= -2C, \,\,[C, H]=D, \,\,
\qquad\qquad\qquad \nonumber\\
\,\,[H,P_{n,i}]&=&-nP_ {n-1,i},\,\,[D,P_{n,i}]=(1 - 2n)P_ {n,i}, \,\,
[C,P_{n,i}]=(1 - n)P_{n+1,i},
\qquad\qquad\qquad \nonumber\\
\,\,[J_{ij}, P_{n,k} ]&=&\delta_{ik}P_{n,j} - \delta_{jk}P_{n,i},  \,\,
[J_{ij}, J_{k\ell}] =
\delta_{ik}J_{j\ell} + \delta_{j \ell}J_{ik} - \delta_{i \ell}J_{jk} -
\delta_{jk}J_{i \ell},\nonumber\\
\,\,[P_{m,i},P_{n,j}]&=& \delta_{i,j} \delta_{m+n,1} (-1)^{m+1} M. \label{th4} 
\end{eqnarray}
Note that the generators $\{ H , D , C \}$ span an $\mathfrak{sl}(2)$ subalgebra, while the generators
$\{P_{n,i},M\}$ span the Heisenberg Lie algebra $\mathfrak{H}_d$. 
The generator $M$ is in the centre of the algebra, and there is also an $\mathfrak{so}(d)$
subalgebra of spatial rotations generated by $\{J_{ij}\}$. 
The structure of the Levi decomposition is therefore 
$$
\mathfrak{sch}(d)=\mathfrak{sl}(2)\oplus \mathfrak{so}(d) \niplus  \mathfrak{H}_d,
$$
and the dimension is $\frac12 d^2+\frac32 d+4$. 
The Beltrametti-Blasi formula \cite{BelBla66} predicts that there will be
$$
\mbox{rank}(\mathfrak{sl}(2))+\mbox{rank}(\mathfrak{so}(d))+1=\left\lfloor \frac{d}{2} \right\rfloor+2
$$ 
functionally independent Casimir operators, also in agreement with \cite{Ca05}. To demonstrate our approach, 
we focus only on searching for Casimir operators up to quartic order.

Various results have already appeared in the literature concerning the Casimir operators of
$\mathfrak{sch}(d)$. Notably, Perroud \cite{Pe77} first gave the
three functionally independent Casimir operators for $d=3$, including the central element $M$, and two
independent quartic Casimir operators were given. In a sense, one of these could be associated with
the $\mathfrak{sl}(2)$ Levi factor, the other relating to the $\mathfrak{so}(d)$ part. For $d=1$,
Aizawa and Dobrev \cite{Ai10} made use of
the form of the $\mathfrak{sl}(2)$ related Casimir operator given by Perroud, and rightly observe
that the central element $M$ may be factored out, essentially leaving a cubic Casimir operator. In the
methods of Campoamor-Stursberg and Low \cite{Ca09} and Quesne \cite{Qu88} are also suitable for
$\mathfrak{sch}(d)$, although the methods have not been specifically applied to this algebra, but as
we have already mentioned, $\mathfrak{sch}(d)$ has been studied in the context of matrix methods of
contructing Casimir operators in \cite{Ca05}. 
 
From the Lie bracket given above, it is clear that one possible way of interpreting the level of the
generators in terms of artificial relative dimensions $a$ and $b$ is as follows: 
\begin{align*}
[P_{0,j}]=ab, \, [P_ {1,j}]=a^{-1}b,\,
[H]={a^{2}} ,
\, 
[C]={{a^{-2}}}, \, [D]=1=[J_{ij}], \,  [M]=b^2,
\end{align*} 
or simply
\begin{align*}
[P_{0,j}]=(1,1), \, [P_ {1,j}]=(-1,1),\,
[H]=(2,0),
\, 
[C]=(-2,0), \, [D]=(0,0)=[J_{ij}], \,  [M]=(0,2).
\end{align*} 
It is at this point that a comment is in order about the dimensional analysis. This step is useful
in simplifying the process of taking linear combinations of terms in steps (B') and (C') of the
modified search algorithm. While there is not a unique choice,\footnote{e.g. we could choose $[X]=1$ for all
$X$, but this would reduce to the original naive algorithm} for Lie algebras with a non-trivial
Levi factor such as $\mathfrak{sch}(d)$, a canonical way could be to make use of the root system of the
semisimple part. In other words, to use a Cartan-Weyl basis for the semisimple part, and then the
corresponding root system will provide a suitable structure for the artificial relative dimensions of these
generators. 
For $\mathfrak{sch}(d)$, the Levi factor is $\mathfrak{sl}(2)\oplus\mathfrak{so}(d).$
The basis $\{ J_{ij}\}$ that we use for the $\mathfrak{so}(d)$ part is clearly not in Cartan-Weyl
form. Our choice of artificial relative dimensions for these $\mathfrak{so}(d)$ generators is clearly trivial, but this
still serves our purpose for determining low order Casimir operators and is computationally
feasible. In fact, the convenience of our current approach is that the number of independent labels
does not grow with $d$. If we were to use a labelling scheme for the artificial relative dimensions
based on the roots of the Levi factor, the number of independent labels would indeed increase by
$\lfloor d/2 \rfloor$. Our
approach fixes the number of labels at two for all values of $d$. 
Further discussion and details of dimensional analysis, root systems and Cartan-Weyl basis in this setting will
feature in forthcoming work.

The differential operator realisation used for this example (see \cite{Do97} for more detail) is the
vector field realisation
\begin{eqnarray*}
P_ {0,j}&=&\frac{\partial} {\partial x_{j}} ,\  \,\,\,\,\,\,\,\  
P_ {1,j}=-t\frac{\partial} {\partial x_{j}}-mx_j ,\   \,\,\,\,\,\,\,\  
M=m ,
\qquad\qquad\qquad \\
H&=&\frac{\partial}{\partial t},\,\,\,\,\,\,\,\ 
D=-2t\frac{\partial}{\partial t} -x_{k}\frac{\partial} {\partial x_{k}}-\frac{1}{2} ,
\qquad\qquad\qquad \\
C&=&t^2\frac{\partial } {\partial t}+t\,x_{k}\frac{\partial}  {\partial x_{k}}+\frac{1}
{2}mx_{k}x_{k}+\frac{t}{2},\,\,\,\,\,\ 
J_{ij}=-x_{i}\frac{\partial} {\partial\ x_{j}}+x_{j}\frac{\partial} {\partial x_{i}},
\end{eqnarray*}
where we use the standard convention of summation over repeated indices. 

We proceed by applying the search algorithm for $\mathfrak{sch}(d)$ to the specific cases where
$d=1,2,3,4$, and then make some comment on
generalisations to the case of arbitrary $d$. Our search algorithm has found Casimir operators with the same
artificial relative dimensions as the central element $M$, i.e. $(0,2)$ and $M^2$
(relative dimension $(0,4)$). In each case we give all expressions in terms of the generators given
in (\ref{Bas}) above.

Throughout, a Casimir operator of order $n$ for $\mathfrak{sch}(d)$ of
relative dimension $(s,t)$ is denoted as
$$
K_{(s,t)}^{(n),d}.
$$
The results are summarised below. We also make reference to the spurious candidate Casimir operators which
arise from step (H').  

\subsubsection{$\mathfrak{sch}(1)$}

As already discussed, for this case the Beltrametti-Blasi formula predicts two Casimir operators for
$\mathfrak{sch}(1)$. One is clearly the central element $M$. Using our algorithm, the search for a
cubic Casimir operator with artificial relative dimension $(0,2)$ found
$$
K_{(0,2)}^{(3),1} = MD^{2}-3MD-4MHC+2P^{2}_{1,1}H+2P^{2}_{0,1}C-2P_{0,1}P_{1,1}D,
$$
using a linear combination of terms up to degree 3. 

We remark that for this example, the method of virtual copies of Campoamor-Stursberg and Low \cite{Ca09}
utilises elements
\begin{align*}
\tilde{H} &= MH - \frac12 P_{0,1}^2,\\
\tilde{C} &= MC-\frac12P_{1,1}^2,\\
\tilde{D} &= MD - \frac12 M - P_{0,1}P_{1,1},
\end{align*}
which satisfy the relations
\begin{equation}
[\tilde{D},\tilde{H}]=2M\tilde{H},\ \ [\tilde{D},\tilde{C}]=-2M\tilde{C},\ \
[\tilde{C},\tilde{H}]=M\tilde{D},
\label{vsl2}
\end{equation}
and thus form a virtual copy of $\mathfrak{sl}(2)$. It is also easily seen that the elements
$\tilde{H},$ $\tilde{C}$ and $\tilde{D}$ commute with $P_{0,1}$ and $P_{1,1}$, and so writing down
the usual quadratic $\mathfrak{sl}(2)$ Casimir operator in terms of $\tilde{H},$ $\tilde{C}$ and $\tilde{D}$ 
will produce a genuine Casimir operator for $\mathfrak{sch}(1)$. In fact, this is somewhat related to
the original approach of Perroud \cite{Pe77}. It is no surprise that the
resulting Casimir operator
\begin{equation}
K = \tilde{D}^2-2\tilde{H}\tilde{C} - 2\tilde{C}\tilde{H}
\label{sl2cas}
\end{equation}
is functionally related to $K_{(0,2)}^{(3),1}$ (presented above) by
$$
MK_{(0,2)}^{(3),1} + K + \frac34M^2 = 0.
$$
The Casimir operator $K$ is the same one used by Aizawa and Dobrev (up to a scalar factor) in
\cite{Ai10}.

\subsubsection{$\mathfrak{sch}(2)$}

For this case the Levi decomposition is 
\begin{equation}
\mathfrak{sl}(2)\oplus \mathfrak{so}(2) \niplus  \mathfrak{H}_2,
\label{Levi2}
\end{equation}
and there are three functionally independent Casimir operators, including $M$. 
Using our algorithm, a quadratic Casimir operator was found with artificial relative dimensions $(0,2)$ as follows:
$$
K_{(0,2)}^{(2),2} = M J_{12} + P_{0,1} P_{1,2} - P_{0,2} P_{1,1}.
$$
A functionally independent cubic Casimir was also found with artificial relative dimensions $(0,2)$:
\begin{align*}
K_{(0,2)}^{(3),2} =& MD^{2}-4MD-4MHC+2P^{2}_{1,1}H+2P^{2}_{1,2}H+2P^{2}_{0,1}C+2P^{2}_{0,2}C
 \\
& \quad
-2P_{0,1}P_{1,1}D-2P_{0,2}P_{1,2}D+MJ^{2}_{12}+2P_{0,1}P_{1,2}J_{12}-2P_{0,2}P_{1,1}J_{12}.
\end{align*}
For this cubic case, the output at step (G') produces two candidate expressions as
\begin{align*}
K_a=& MD^{2}- 4MD-4MHC+2P^{2}_{1,1}H+2P^{2}_{1,2}H+2P^{2}_{0,1}C
 \\
& \quad +2P^{2}_{0,2}C-2P_{0,1}P_{1,1}D - 2P_{0,2}P_{1,2}D-MJ^{2}_{12},\\
 K_b=&M J_{12}^2 + P_{0,1} P_{1,2} J_{12} - P_{0,2} P_{1,1} J_{12}.
\end{align*}
It turns out neither of these are Casimir operators, but behave as such when restricted to the
realisation. Taking a linear combination results in 
$$
K_a+2K_b=K_{(0,2)}^{(3),2}.
$$

We note that $K_{(0,2)}^{(2),2}$ is a Casimir operator related only to the $\mathfrak{so}(2)$ part of
the Levi factor in (\ref{Levi2}). 

By contrast, the method of virtual copies \cite{Ca09} in this case relates to elements
\begin{align}
\tilde{H} &= MH - \frac12 P_{0,1}^2-\frac12 P_{0,2}^2, \label{s1}\\
\tilde{C} &= MC-\frac12P_{1,1}^2-\frac12 P_{1,2}^2,\label{s2}\\
\tilde{D} &= MD - M - P_{0,1}P_{1,1} - P_{0,2}P_{1,2}, \label{s3}
\end{align}
which form a virtual copy of $\mathfrak{sl}(2)$, i.e. satisfy the relations (\ref{vsl2}). These
three generators, along with 
$$
\tilde{J}_{12} = M J_{12} + P_{0,1} P_{1,2} - P_{0,2} P_{1,1},
$$
which forms a virtual copy of $\mathfrak{so}(2)$ (and is precisely $K_{(0,2)}^{(2),2}$), 
all commute with the $P_{n,j}$. 
By this approach, there is one Casimir operator, $K$, associated with
$\mathfrak{sl}(2)$, given by the expression (\ref{sl2cas}), but using the generators given in
(\ref{s1})-(\ref{s3}). 
Explicitly, this gives
\begin{align*}
K =& M^2D^2 - 4M^2D-4M^2HC+2MP^{2}_{1,1}H+2MP^{2}_{1,2}H+2MP^{2}_{0,1}C+2MP^{2}_{0,2}C
 \\
& \quad
-2MP_{0,1}P_{1,1}D-2MP_{0,2}P_{1,2}D + MP_{0,1}P_{1,1} + MP_{0,2} P_{1,2} +
2P_{0,1}P_{0,2}P_{1,1}P_{1,2} \\
& \quad - P_{0,1}^2P_{1,2}^2-P_{0,2}^2P_{1,1}^2 - M^2.
\end{align*}
We then have the functional relation
$$
K + \left(K_{(0,2)}^{(2),2}\right)^2 +M^2 = MK_{(0,2)}^{(3),2}.
$$
From these two examples of $\mathfrak{sch}(1)$ and $\mathfrak{sch}(2)$, we are beginning to have a
picture of what our algorithm can achieve. In particular, we see that our algorithm is able to
produce low order Casimir operators.

\subsubsection{$\mathfrak{sch}(3)$}

For the case $d=3$, we expect three functionally independent Casimir operators, as in the case $d=2$. 
Searching for a cubic Casimir operator of artificial relative dimensions $(0,2)$ -- the same dimensions as
the central element $M$ -- yields the following four spurious candidates:
\begin{align}
K_a=&-5 MD+MD^{2}-4MHC+2P^{2}_{1,1}H+2P^{2}_{1,2}H+2P^{2}_{1,3}H 
\nonumber\\
& \quad
+2P^{2}_{0,1}C+2P^{2}_{0,2}C+2P^{2}_{0,3}C-2P_{0,1}P_{1,1}D-2P_{0,2}P_{1,2}D-2P_{0,3}P_{1,3}D
\nonumber\\
& \quad -MJ^{2}_{12}-MJ^{2}_{13}-MJ^{2}_{23}, \label{spcand1a}\\  
K_b =& MJ^{2}_{12}-P_ {0,2} P_{1,1}J_{12}+P_ {0,1} P_{1,2} J_{12}, \label{spcand1b}\\  
K_c =& MJ^{2}_{13}-P_ {0,3} P_{1,1}J_{13}+P_ {0,1} P_{1,3}J_{13}, \label{spcand1c}\\    
K_d =& MJ^{2}_{23}-P_ {0,3} P_{1,2}J_{23}+P_ {0,2} P_{1,3}J_{23}. \label{spcand1d}
\end{align}
These spurious candidates behave as a Casimir operator only when restricted to the realisation. 
It is easily checked that taking the following linear combination produces a genuine cubic Casimir
operator:
$$
K_a+2( K_b+ K_c+K_d ) = K_{(0,2)}^{(3),3}. 
$$
Explicitly we have
\begin{align*}
K_{(0,2)}^{(3),3} =& MD^{2} -5 MD-4MHC+2P^{2}_{1,1}H+2P^{2}_{1,2}H+2P^{2}_{1,3}H \\
& \quad +2P^{2}_{0,1}C+2P^{2}_{0,2}C+2P^{2}_{0,3}C-2P_{0,1}P_{1,1}D-2P_{0,2}P_{1,2}D-2P_{0,3}P_{1,3}D \\
& \quad +MJ^{2}_{12}+MJ^{2}_{13}+MJ^{2}_{23}-2P_ {0,2} P_{1,1}J_{12}+2P_ {0,1} P_{1,2} J_{12} -2P_
{0,3} P_{1,1}J_{13}\\
& \quad +2P_ {0,1} P_{1,3}J_{13}-2P_ {0,3} P_{1,2}J_{23}2+P_ {0,2} P_{1,3}J_{23}.
\end{align*}
Next, searching for a quartic Casimir operator of artificial relative dimensions $(0,4)$ (equivalent
to $M^2$), leads to the following nine spurious candidates, all of which behave as Casimir operators only when
restricted to the realisation:
\begin{align*}
K_a =& M^2 J^{2}_{12}-M P_{0,1}P_{1,1}+M P_{0,2}P_{1,2}+ 2M P_{0,1}P_{1,2}J_{12}-P^{2}_{0,2}P^{2}_{1,1}+P^{2}_{0,1}P^{2}_{1,2}, \\
K_b =& M^2 J^{2}_{12}+M P_{0,1}P_{1,1}+M P_{0,2}P_{1,2}-P^{2}_{0,1}P^{2}_{1,2}-P^{2}_{0,2}P^{2}_{1,1}+2P_{0,1}P_{0,2} P_{1,1}P_{1,2}, \\  
K_c =& -M^2 J^{2}_{12}-M P_{0,1}P_{1,1}+M P_{0,2}P_{1,2}+ 2M P_{0,2}P_{1,1}J_{12}-P^{2}_{0,2}P^{2}_{1,1}+P^{2}_{0,1}P^{2}_{1,2}, \\ 
K_d =& M^2 J^{2}_{13}-M P_{0,1}P_{1,1}+M P_{0,3}P_{1,3}+ 2M P_{0,1}P_{1,3}J_{13}-P^{2}_{0,3}P^{2}_{1,1}+P^{2}_{0,1}P^{2}_{1,3}, \\
K_e =& M^2 J^{2}_{13}+M P_{0,1}P_{1,1}+M P_{0,3}P_{1,3}-P^{2}_{0,1}P^{2}_{1,3}-P^{2}_{0,3}P^{2}_{1,1}+2P_{0,1}P_{0,3} P_{1,1}P_{1,3}, \\  
K_f =& -M^2 J^{2}_{13}-M P_{0,1}P_{1,1}+M P_{0,3}P_{1,3}+ 2M P_{0,3}P_{1,1}J_{13}-P^{2}_{0,3}P^{2}_{1,1}+P^{2}_{0,1}P^{2}_{1,3}, \\
K_g =& M^2 J^{2}_{23}-M P_{0,2}P_{1,2}+M P_{0,3}P_{1,3}+ 2M P_{0,2}P_{1,3}J_{23}-P^{2}_{0,3}P^{2}_{1,2}+P^{2}_{0,2}P^{2}_{1,3}, \\
K_h =& M^2 J^{2}_{23}+M P_{0,2}P_{1,2}+M P_{0,3}P_{1,3}-P^{2}_{0,2}P^{2}_{1,3}-P^{2}_{0,3}P^{2}_{1,2}+2P_{0,2}P_{0,3} P_{1,2}P_{1,3} \\  
K_i =& -M^2 J^{2}_{23}-M P_{0,2}P_{1,2}+M P_{0,3}P_{1,3}+ 2M P_{0,3}P_{1,2}J_{23}-P^{2}_{0,3}P^{2}_{1,2}+P^{2}_{0,2}P^{2}_{1,3}. 
\end{align*}
The following linear combination produces a genuine quartic Casimir operator:
$$
K_a-( K_b+ K_c)+K_d-( K_e+ K_f)+K_g-( K_h+ K_i) = K_{(0,4)}^{(4),3},
$$
which is expressed explicitly as
\begin{align*}
K_{(0,4)}^{(4),3} =&-2M( P_{0,1}P_{1,1}+  P_{0,2}P_{1,2}+  P_{0,3}P_{1,3})+M^2( J^{2}_{12}+J^{2}_{13} +J^{2}_{23} )+2M ( P_{0,1}P_{1,2}- P_{0,2}P_{1,1}) J_{12} \\
&\quad +2M ( P_{0,1}P_{1,3}- P_{0,3}P_{1,1}) J_{13}+2M ( P_{0,2}P_{1,3}- P_{0,3}P_{1,2}) J_{23}+ P^{2}_{0,1}P^{2}_{1,2}+ P^{2}_{0,2}P^{2}_{1,1} \\
&\quad + P^{2}_{0,1}P^{2}_{1,3}+ P^{2}_{0,3}P^{2}_{1,1}+ P^{2}_{0,2}P^{2}_{1,3}+ P^{2}_{0,3}P^{2}_{1,2}-2P_{0,1}P_{0,2} P_{1,1}P_{1,2}-2P_{0,1}P_{0,3} P_{1,1}P_{1,3} \\
&\quad -2P_{0,2}P_{0,3} P_{1,2}P_{1,3}. 
\end{align*}

The method of virtual copies \cite{Ca09} in this case relates to elements
\begin{align}
\tilde{H} &= MH - \frac12 P_{0,1}^2-\frac12 P_{0,2}^2-\frac12 P_{0,3}^2, \label{s13}\\
\tilde{C} &= MC-\frac12P_{1,1}^2-\frac12 P_{1,2}^2-\frac12 P_{1,3}^2,\label{s23}\\
\tilde{D} &= MD - \frac32M - P_{0,1}P_{1,1} - P_{0,2}P_{1,2}- P_{0,3}P_{1,3}, \label{s33}
\end{align}
which form a virtual copy of $\mathfrak{sl}(2)$, i.e. satisfy the relations (\ref{vsl2}). Also, 
the elements 
\begin{align*}
\tilde{J}_{12} =& M J_{12} + P_{0,1} P_{1,2} - P_{0,2} P_{1,1},\\
\tilde{J}_{13} =& M J_{13} + P_{0,1} P_{1,3} - P_{0,3} P_{1,1},\\
\tilde{J}_{23} =& M J_{23} + P_{0,2} P_{1,3} - P_{0,3} P_{1,2},
\end{align*}
form a virtual copy of $\mathfrak{so}(3)$ in the sense that they satisfy the relations
$$
[\tilde{J}_{12},\tilde{J}_{13}] = M\tilde{J}_{23},\ \ [\tilde{J}_{12},\tilde{J}_{23}] =
-M\tilde{J}_{13},\ \ [\tilde{J}_{13},\tilde{J}_{23}] = M\tilde{J}_{12}.
$$ 
Here, all elements $\{ \tilde{H},\tilde{C},\tilde{D}, \tilde{J}_{12},\tilde{J}_{13},\tilde{J}_{23} \}$ commute with the $P_{n,j}$.
As in the $d=2$ case, there is one Casimir operator associated with
$\mathfrak{sl}(2)$, given by the expression (\ref{sl2cas}), but using the generators given in
(\ref{s13})-(\ref{s33}). There is also another Casimir operator which corresponds to the quadratic Casimir operators
of $\mathfrak{so}(3)$, $\tilde{J}_{12}^2+\tilde{J}_{13}^2+\tilde{J}_{23}^2$ which in this case
coincides precisely with $K_{(0,4)}^{(4),3}$ given above. Moreover, we find the function relation
$$
K = MK_{(0,2)}^{(3),3} - K_{(0,4)}^{(4),3},
$$
where $K$ is the Casimir operator of (\ref{sl2cas}) related to the generators of the virtual
$\mathfrak{sl}(2)$ given in (\ref{s13})-(\ref{s33}).

\subsubsection{$\mathfrak{sch}(4)$}

For this case, we expect four functionally independent Casimir operators. We note that the Levi
decomposition is
$$
\mathfrak{sl}(2)\oplus \mathfrak{so}(4) \niplus  \mathfrak{H}_4,
$$
and in this special case we also have that $\mathfrak{so}(4) \cong \mathfrak{so}(3)\oplus
\mathfrak{so}(3)$ as a Lie algebra.

In applying our algorithm, we actually found two functionally independent cubic Casimir operators of artificial relative dimensions
$(0,2)$. One is given by
\begin{align*}
K_{(0,2)}^{(3),4} =& MD^{2} -6 MD-4MHC+2(P^{2}_{1,1}+P^{2}_{1,2}+P^{2}_{1,3}+P^{2}_{1,4})H+2(P^{2}_{0,1}+P^{2}_{0,2}+P^{2}_{0,3}+ P^{2}_{0,4})C
\\
& \quad -2(P_{0,1}P_{1,1}+P_{0,2}P_{1,2}+P_{0,3}P_{1,3}+P_{0,4}P_{1,4})D
+M (J^{2}_{12}+J^{2}_{13}+J^{2}_{14}+J^{2}_{23}+J^{2}_{24}+J^{2}_{34})
\\
& \quad -2(P_{0,2}P_{1,1} -P_{0,1}P_{1,2})J_{12}-2(P_{0,3}P_{1,1} -P_{0,1}P_{1,3})J_{13}
-2(P_{0,4}P_{1,1} -P_{0,1}P_{1,4})J_{14}\\
& \quad
-2(P_{0,3}P_{1,2}-P_{0,2}P_{1,3})J_{23}-2(P_{0,4}P_{1,2}-P_{0,2}P_{1,4})J_{24}-2(P_{0,4}P_{1,3}-P_{0,3}P_{1,4})J_{34}.
\end{align*}
The second, denoted $\bar{K}_{(0,2)}^{(3)}$, is given as
\begin{align*}
\bar{K}_{(0,2)}^{(3)} =& M( J_{12}J_{34}-J_{13}J_{24}+J_{14}J_{23} )+ ( P_{0,1}P_{1,2}- P_{0,2}P_{1,1}) J_{34}+ ( P_{0,3}P_{1,1}- P_{0,1}P_{1,3}) J_{24}
\\
& \quad + ( P_{0,2}P_{1,3}- P_{0,3}P_{1,2}) J_{14}+ ( P_{0,1}P_{1,4}- P_{0,4}P_{1,1}) J_{23}+ ( P_{0,4}P_{1,2}- P_{0,2}P_{1,4}) J_{13}\\
&\quad + ( P_{0,3}P_{1,4}- P_{0,4}P_{1,3}) J_{12}.
\end{align*}
When applying our algorithm to search for cubic Casimir operators of artificial relative dimension
$(0,2)$, 17 spurious candidates are returned (we will not give the list here). 
When taking a linear combination of these, we find that there are the two functionally independent solutions.
A genuine quartic Casimir operator was also found of artificial relative dimension $(0,4)$. It was
the result of a linear combination of 18 spurious candidates (not given here), and turns out to be 
\begin{align*}
K_{(0,4)}^{(4),4} =& -3M( P_{0,1}P_{1,1}+  P_{0,2}P_{1,2}+  P_{0,3}P_{1,3}+  P_{0,4}P_{1,4})+M^2( J^{2}_{12}+J^{2}_{13}+J^{2}_{14} +J^{2}_{23} +J^{2}_{24}+J^{2}_{34})\\
&\quad
+2M ( P_{0,1}P_{1,2}- P_{0,2}P_{1,1}) J_{12}+2M ( P_{0,1}P_{1,3}- P_{0,3}P_{1,1}) J_{13}+2M ( P_{0,1}P_{1,4}- P_{0,4}P_{1,1}) J_{14}\\
&\quad
+2M ( P_{0,2}P_{1,3}- P_{0,3}P_{1,2}) J_{23}+2M ( P_{0,2}P_{1,4}- P_{0,4}P_{1,2}) J_{24}+2M ( P_{0,3}P_{1,4}- P_{0,4}P_{1,3}) J_{34} \\
& \quad
+ (P^{2}_{0,1}P^{2}_{1,2}+ P^{2}_{0,2}P^{2}_{1,1}) + (P^{2}_{0,1}P^{2}_{1,3}+ P^{2}_{0,3}P^{2}_{1,1})+ (P^{2}_{0,1}P^{2}_{1,4}+ P^{2}_{0,4}P^{2}_{1,1}) \\
& \quad
 + (P^{2}_{0,2}P^{2}_{1,3}+ P^{2}_{0,3}P^{2}_{1,2})+ (P^{2}_{0,2}P^{2}_{1,4}+ P^{2}_{0,4}P^{2}_{1,2})+ (P^{2}_{0,3}P^{2}_{1,4}+ P^{2}_{0,4}P^{2}_{1,3}) \\
&\quad
- 2P_{0,1}P_{0,2} P_{1,1}P_{1,2} - 2P_{0,1}P_{0,3} P_{1,1}P_{1,3}
- 2P_{0,1}P_{0,4} P_{1,1}P_{1,4} - 2P_{0,2}P_{0,3} P_{1,2}P_{1,3} \\
& \quad
- 2P_{0,2}P_{0,4} P_{1,2}P_{1,4} - 2P_{0,3}P_{0,4} P_{1,3}P_{1,4}
\end{align*}

Once again, we provide a comparison to the method of virtual copies \cite{Ca09}, which makes use of elements
\begin{align}
\tilde{H} &= MH - \frac12 (P_{0,1}^2+P_{0,2}^2+ P_{0,3}^2), \label{s14}\\
\tilde{C} &= MC-\frac12(P_{1,1}^2+P_{1,2}^2+P_{1,3}^2),\label{s24}\\
\tilde{D} &= MD - 2M - P_{0,1}P_{1,1} - P_{0,2}P_{1,2}- P_{0,3}P_{1,3}, \label{s34}
\end{align}
which satisfy the relations (\ref{vsl2}), and thus constitute a virtual copy of $\mathfrak{sl}(2)$. Also, 
the elements 
$$
\tilde{J}_{ij} = M J_{ij} + P_{0,i} P_{1,j} - P_{0,j} P_{1,i},\ \ 1\leq i<j\leq 4,
$$
form a virtual copy of $\mathfrak{so}(4)$ in the sense that they satisfy the relations
\begin{equation}
[\tilde{J}_{ij},\tilde{J}_{k\ell}] = 
\delta_{ik}M\tilde{J}_{j\ell} + \delta_{j \ell}M\tilde{J}_{ik} - \delta_{i \ell}M\tilde{J}_{jk} -
\delta_{jk}M\tilde{J}_{i \ell}.
\label{vsod}
\end{equation} 
As already mentioned, $\mathfrak{so}(4)$ is a special case in that $\mathfrak{so}(4) \cong \mathfrak{so}(3)\oplus
\mathfrak{so}(3)$ as a Lie algebra. It is easily verified that two 
Casimir operators for $\mathfrak{sch}(4)$ related to $\mathfrak{so}(4)$ are associated with the two
well-known quadratic expressions for the $\mathfrak{so}(4)$ Casimir operators, namely
\begin{align}
K_1 =&
\tilde{J}_{12}^2+\tilde{J}_{13}^2+\tilde{J}_{14}^2+\tilde{J}_{23}^2+\tilde{J}_{24}^2+\tilde{J}_{34}^2,
\label{so41} \\
K_2 =& \tilde{J}_{12}\tilde{J}_{34}-\tilde{J}_{13}\tilde{J}_{24}+\tilde{J}_{14}\tilde{J}_{23}.
\label{so42}
\end{align}
By direct calculation, and analogous to the $d=3$ case, we see that 
$$
K = MK_{(0,2)}^{(3),4} - K_{(0,4)}^{(4),4},
$$
where $K=\tilde{D}^2-2\tilde{H}\tilde{C} - 2\tilde{C}\tilde{H}$ is associated with the quadratic
Casimir from the virtual copy of $\mathfrak{sl}(2)$. It also turns that the Casimir operators of
(\ref{so41}) and (\ref{so42}) above coincide with the other two produced by our algorithm, i.e.
\begin{align*}
K_{(0,4)}^{(4),4} =& K_1,\\
M\bar{K}_{(0,2)}^{(3),4} =& K_2.
\end{align*} 

\subsubsection{$\mathfrak{sch}(d)$}

For the case of general $d$, we do not give a full set of functionally independent Casimir operators since
the expressions are indeed cumbersome, but are able to make some statements about the general form
of some of the Casimir operators we have already seen.

Firstly, for all $d$, there exists a cubic Casimir operator of artificial relative dimension $(0,2)$
(in terms of the consistent scheme used in this paper) given by
\begin{align*}
K_{(0,2)}^{(3),d} =& MD^2 - (d+2)MD -4HC + 2\left( \sum_{i=1}^d\left( P_{1,i}^2H+P_{0,i}^2C -
P_{0,i}P_{1,i}D \right) \right) \\
& \quad +\sum_{i=1}^{d-1}\sum_{j=i+1}^d\left( MJ_{ij} + 2(P_{0,i}P_{1,j} - P_{0,j}P_{1,j})
\right)J_{ij}.
\end{align*}
Note that in the case $d=1$, the double sum in the last term does not appear.

Secondly, for the case $d>2$, there is a quartic Casimir operator of artificial relative dimension
$(0,4)$ given by
\begin{align*}
K_{(0,4)}^{(4),d} =& \sum_{i=1}^{d-1}\sum_{j=i+1}^d\left( M^2J_{ij}^2+2M(P_{0,i}P_{1,j} -
P_{0,j}P_{1,i})J_{ij} + P_{0,i}^2P_{1,j}^2 + P_{0,j}^2P_{1,i}^2 -2P_{0,i}P_{1,i}P_{0,j}P_{1,j}
\right) \\
& \quad -(d-1)M\sum_{i=1}^dP_{0,i}P_{1,i}.
\end{align*}
Both Casimir operators presented may be verified by direct calculation (using the general
commutation relations (\ref{th4})) that they do indeed commute with all generators. 

Regarding the method of virtual copies \cite{Ca09}, it can be seen that
\begin{align*}
\tilde{H} &= MH - \frac12 \sum_{i=1}^d P_{0,i}^2,\\
\tilde{C} &= MC-\frac12 \sum_{i=1}^d P_{1,i}^2,\\
\tilde{D} &= MD - \frac{d}{2}M - \sum_{i=1}^dP_{0,i}P_{1,i}
\end{align*}
produce a virtual copy of $\mathfrak{sl}(2)$, satisfying relations (\ref{vsl2}). Also, the elements
\begin{equation}
\tilde{J}_{ij} = M J_{ij} + P_{0,i} P_{1,j} - P_{0,j} P_{1,i},\ \ 1\leq i<j\leq d,
\label{vsodels}
\end{equation}
can be shown by tedious calculation to produce a virtual copy of $\mathfrak{so}(d)$, satisfying
relations (\ref{vsod}). A way of determining the higher order Casimir operators associated with
the virtual copy of $\mathfrak{so}(d)$ is to make use of the methods of Gruber and O'Rafeartaigh \cite{Gr64}.
To this end, we arrange the elements $\tilde{J}$ into a matrix as follows:
$$
\tilde{J} =
\left(
\begin{array}{ccccccc}
0              & -\tilde{J}_{12} & -\tilde{J}_{13} & -\tilde{J}_{14} & \cdots & -\tilde{J}_{1 (d-1)} & -\tilde{J}_{1d}\\
\tilde{J}_{12} & 0               & -\tilde{J}_{23} & -\tilde{J}_{24} & \cdots & -\tilde{J}_{2 (d-1)} & -\tilde{J}_{2d}\\
\tilde{J}_{13} & \tilde{J}_{23}  & 0               & -\tilde{J}_{34} & \cdots & -\tilde{J}_{3 (d-1)} & -\tilde{J}_{3d}\\
\vdots         &                 &                 &           &   \ddots     &                      & \vdots \\
\tilde{J}_{1(d-1)} &              &                 &                 &        & 0                 &-\tilde{J}_{(d-1)d} \\
\tilde{J}_{1d} & \cdots          &                 &                 & \cdots & \tilde{J}_{(d-1)d} & 0
\end{array}
\right)
$$
and define powers of this matrix using regular matrix multiplication, but taking care of the order of
products of the non-commuting entries. 
The following expressions are known to correspond to functionally independent Casimir operators of
$\mathfrak{so}(d)$:
$$
I_{2r} = \mbox{tr}\left( \tilde{J}^{2r} \right), \ \ r=1,2,\ldots,\left\lfloor \frac{d}{2} \right\rfloor.
$$
In the case of $\mathfrak{sch}(d),$ we insert the expressions in (\ref{vsodels}) for the entries of
the matrix. Generally the resulting expression will give a Casimir operator for $\mathfrak{sch}(d)$
of order $4r$ of artificial relative dimension $(0,4r)$.

It is worth remarking that in the case of $d=2$, the resulting Casimir operator from this approach is simply
$\tilde{J}_{12}^2,$ but for that case it turns out that $\tilde{J}_{12}$ is actually a Casimir
operator. Also, for $d=4,$ $r=1$ in the above formula will produce $K_{(0,4)}^{(4),4}$ given above,
and $r=2$ will produce a Casimir operator of order 8 which will be functionally dependent on those
already given above. Apart from these exceptional cases, for all other values of $d$, this approach
will produce the required number of functionally independent Casimir operators.

\section{Conclusion}

In this paper, we have presented a simple search algorithm for the construction of polynomial
Casimir operators of non-semisimple Lie algebras. The algorithm makes use of differential operator
realisations, and simplifies calculations by using artificial relative dimensions, a construct not
dissimilar to root systems for simple Lie algebras. 

One common attribute featured in the two key examples of model filiform Lie algebras and
Schr\"odinger Lie algebras is that our search algorithm seems to be effective in finding a lowest
order set (in a loose sense) of functionally independent Casimir operators. This is evident from the
output for the example of model filiform Lie algebras, where our algorithm produced quadratic and
cubic Casimir operators only, in contrast to the output of the infinitesimal approach.
When applied to the Schr\"{o}dinger Lie algebras corresponding to low underlying spatial dimension,
we find that our search algorithm returns a particular linear combination of Casimir operators found
using the methods of \cite{Qu88,Ca09} that results in low order operators.

There are various generalisations possible to structures such as Lie superalgebras \cite{Kac77} or finite 
polynomial algebras that play a role in context of superintegrable systems \cite{Zhe1} for which the construction of Casimir
operators remains unexplored and Casimir operators are know only for limited classes of such
algebraic structures.

In future work we intend to use our algorithm to determine Casimir operators for the conformal
Galilei algebras for $\ell>\frac12$, and to investigate its usefulness in cases of contractions of
Casimir operators (see for example \cite{WW08}).

Also of interest would be to further develop our use of dimensional analysis into a solid theory for
non semisimple Lie algebras that generalises the notion of root system for simple Lie algebras.

{\bf Acknowledgements:}

The research of FA is supported by Prince Sattam Bin Abdulaziz University. IM is supported
by the Australian Research Council Discovery Grant DP160101376. PSI is supported by the Australian Research Council
through Discovery Project DP150101294.

\end{document}